\documentclass[preprint,aps,showpacs,nofootinbib,nobibnotes]{revtex4}%
\usepackage{amsmath}
\usepackage{amsfonts}
\usepackage{amssymb}%
\usepackage{graphicx}
\providecommand{\U}[1]{\protect\rule{.1in}{.1in}}
\begin{document}
\title{$Z_{3}$ Symmetry and Neutrino Mixing in Type II Seesaw}
\author{Bo Hu$^{1,2}$}
\email{bohu@ncu.edu.cn}
\author{Feng Wu$^{2}$}
\email{fengwu@itp.ac.cn}
\author{Yue-Liang Wu$^{2}$}
\email{ylwu@itp.ac.cn}
\affiliation{${}^{1}$ Department of Physics, Nanchang University, 330031, China}
\affiliation{${}^{2}$ Kavli Institute for Theoretical Physics China (KITPC), Institute of
Theoretical Physics, Chinese Academy of Sciences, P.O. Box 2735, Beijing
100080, China}

\begin{abstract}
Neutrino mixing matrix satisfying the current experimental data can be well
described by the HPS tri-bimaximal mixing matrix. We propose that its origin
can be understood within the seesaw framework by a hidden condition on the
mass matrix of heavy right-handed neutrinos under the transformation of the
Abelian finite group $Z_{3}$ on the flavor basis. Ignoring CP phases, we show
that it can lead to the generic form of the effective light neutrino mass
matrix from which the HPS mixing matrix appears naturally, as well as an
expeimentally allowed non-zero $\sin\theta_{13}$. We show that the model based
on our proposal is in good agreement with the current experimental data.

\end{abstract}

\pacs{14.60.Pq, 11.30.Hv}
\maketitle


In the standard model (SM), the neutrinos are massless. The results of
neutrino oscillation experiments indicate that neutrinos are massive. Due to
the smallness of neutrino masses, the mechanism in the SM which gives quark
and charged lepton masses is unnatural for neutrinos. Therefore, the observed
neutrino oscillations are considered to be the first convincing evidences of
new physics beyond the SM and have been discussed extensively in literatures
(for recent reviews, see \cite{MohapatraReview, ValleReview}).

Besides neutrino masses, the global fit of current experimental data shows
that, unlike the mixing angles in the quark sector, two of the three mixing
angles are large and one of them might be maximal. As a matter of
fact,\ \ $30\mbox{}^{\circ}<\theta_{\mathrm{sol}}<38\mbox{}^{\circ}$,
$36\mbox{}^{\circ}<\theta_{\mathrm{atm}}<54\mbox{}^{\circ}$, and
$\theta_{\mathrm{CHOOZ}}<10\mbox{}^{\circ}$ at the 99\% confidence level
\cite{Strumia:2006db}. To understand this peculiar property is also an
interesting theoretical issue. In fact, these mixing angles can be well
described by the HPS mixing matrix \cite{HPS} where ${\sin}^{2}\theta
_{\mathrm{sol}}=\frac{1}{3}$,\ \ ${\sin}^{2}\theta_{\mathrm{atm}}=\frac{1}{2}%
$, and $\theta_{\mathrm{CHOOZ}}=0$. The HPS mixing matrix can be considered as
the lowest order approximation. Efforts on revealing its orgin may help
understanding not only the neutrino physics, but also the physics beyond the
SM, such as the new symmetries at high energy scales.

With the assumption of the existence of\ \ right-handed neutrinos, seesaw
mechanism provides a simple way to understand the smallness of neutrino masses
and has long been considered as the leading candidate of neutrino mass
generating mechanism. However, by its own seesaw mechanism cannot explain the
observed neutrino mixing pattern.

In many models with right-handed neutrinos, e.g. $\operatorname{SO}(10)$ or
${\operatorname{SU}(2)}_{L}\times{\operatorname{SU}(2)}_{R}\times{U(1)}_{B-L}$
based models, effective light neutrino mass matrix is given by Type II seesaw
relation \cite{seesaw2}
\begin{equation}
M_{\nu}=M_{L}-M_{\nu}^{D}M_{R}^{-1}{\left(  M_{\nu}^{D}\right)  }^{T}.
\label{TypeII}%
\end{equation}

\noindent where $M_{L}$, $M_{R}$ are the majorana mass matrices for
left-handed and right-handed neutrinos and $M_{\nu}^{D}$ the Dirac mass
matrix. Given that the neutrino mass is generated by type II seesaw, as shown
in (\ref{TypeII}), the observed neutrino mixing can provide important
information about the structure of $M_{\nu}$ and thus the physics behind
$M_{L}$, $M_{R}$ and $M_{\nu}^{D}$.

In this Letter we make the natural assumptions that there are three Majorana
neutrinos and consider the case where the flavour symmetry is only broken in
$M_{R}$ sector. Currently none of the CP violating phases has been observed.
In the following discussion, we assume vanishing CP phases and focus on the
mixing pattern. The case with non-vanishing CP phases will be discussed elsewhere.

It is natural to expect that symmetries can lead to specific neutrino mass
matrix. This idea has been pursued in many works. In particular, discrete
symmetries including $S_{3}$ (e.g. \cite{HabaS3}), $S_{4}$ (e.g.
\cite{HagedornS4}), $A_{4}$ (e.g. \cite{MaA4}) etc., have been discussed
extensively in literatures (for recent review, please see
\cite{MohapatraReview} and references therein). Also appropriate flavour
symmetries can also lead to desired neutrino mixing. Without the CP violating
phases, there are six free parameters in $M_{R}$ in the second term of
(\ref{TypeII}), the effective neutrino mass matrix in Type I see-saw. In this
letter, we propose a hidden condition on $M_{R}$ under the transformation of
the Abelian finite group $Z_{3}$ on the flavor basis, which will reduce the
independent parameters down to three. We then use the resultant mass matrix to
explain the observed mixing pattern.

First consider a finite group $G$. Each element $U_{i}$ of $G$ satisfies
$U_{i}^{n_{i}}=1$ for some non-zero integer $n_{i}$. Under an unitary
transformation of $G$ on the flavor basis $\nu_{f}=(\nu_{e},\nu_{\mu}%
,\nu_{\tau})$, we propose that for each $U_{i}$ belongs to $G$, the mass
matrix $M_{R}$ in the new basis satisfies
\begin{equation}
U{}_{i}M_{R}U_{i}^{T}=U_{i}^{^{\prime}}M_{R}. \label{MRR}%
\end{equation}

\noindent We show below that $U_{i}^{^{\prime}}$ is strongly constrained and
any choice of $U_{i}^{^{\prime}}$ satisfying the constraint will further
restrict the possible form of $M_{R}$. In particular, we show that if the
finite group $G$ is chosen to be $Z_{3}$, $U_{i}^{^{\prime}}=U_{i}^{2}$ will
lead to a phenomenologically interesting $M_{\nu}$ and thus provides a
possible origin of the observed neutrino mixing angles.

To see that $U_{i}^{^{\prime}}$ cannot be arbitrary, consider that%
\begin{align}
&  M_{R}={\left(  U_{i}\right)  }^{n_{i}}{M_{R}(U_{i}^{T})}^{n_{i}}={\left(
U_{i}\right)  }^{n_{i}-1}U_{i}^{^{\prime}}{M_{R}(U_{i}^{T})}^{n_{i}%
-1}\nonumber\\
&  ={\left(  U_{i}\right)  }^{n_{i}-1}U_{i}^{^{\prime}}(U_{i}^{\dagger}%
U_{i}^{^{\prime}}){M_{R}(U_{i}^{T})}^{n_{i}-2}=..........\nonumber\\
&  ={\left(  U_{i}\right)  }^{n_{i}-1}{U_{i}^{^{\prime}}(U_{i}^{\dagger}%
U_{i}^{^{\prime}})}^{n_{i}-1}M_{R}={\left(  U_{i}^{\dagger}U_{i}^{^{\prime}%
}\right)  }^{n_{i}}M_{R}. \label{MEM}%
\end{align}

\noindent From (\ref{MEM}) we find that (\ref{MRR}) requires ${(U_{i}%
^{\dagger}U_{i}^{^{\prime}})}^{n_{i}}=1$.\ \ Consequently, we obtain
$U_{i}^{^{\prime}}=e^{\mathrm{i2\pi m}/n_{i}}U_{i}^{k}$ with $m$ some integer
and $k=0,1,\ldots,n_{i}-1$. Note that $m=0$ when $U_{i}$ and $M_{R}$ are real.
Moreover, $k$ can be different for different group element $U_{i}$. Based on
simplicity, we assume $k$ is universal for all group elements. It is obvious
that ${(U_{i}^{\dagger}U_{i}^{^{\prime}})}^{n_{i}}=1$ is only a necessary
condition for (\ref{MRR}) to be held. Given $U_{i}^{^{\prime}}$, (\ref{MRR})
will restrict the form of $M_{R}$. In general, different choice of $k$ will
lead to different $M_{R}$. We show this below in the case where the finite
group $G$ is the cyclic group $Z_{3}$.\ \ 

The group $Z_{3}$ contains only 3 elements, thus $n_{i}\leq3$. Therefore, the
only possible choices for $U_{i}^{^{\prime}}$ are $U_{i}^{^{\prime}}=I$, or
$U_{i}^{^{\prime}}=U_{i}$, or $U_{i}^{^{\prime}}=U_{i}^{2}$. The first choice,
demanding $M_{R}$ to be invariant under $Z_{3}$ on the flavour basis, leads to
an unrealistic mass matrix with $\nu$${}_{e}-\nu_{\mu}-\nu_{\tau}$ symmetry.
Another choice that one might think interesting is the case where
$U_{i}^{^{\prime}}=U_{i}$. One of the necessary conditions in this case
requires $M_{R}$ to be non-invertible so that at least one of the mass
eigenvalues is zero. For the cyclic group $Z_{3}$, the symmetric mass matrix
$M_{R}$ turns out to be democratic in this case and there is only one non-zero
eigenvalue. We won't pursue these in this Letter.

In the following we focus on the case $U_{i}^{\prime}=U_{i}^{2}$. $M_{R}$
built in this way will give interesting phenomenology. In fact, the resultant
$M_{R}$ can be expressed as linear combinations of elements in one of the two
cosets of $Z_{3}$ in the non-Abelian symmetric group $S_{3}$. Our bottom-up
approach ends up with the proposal that under some finite group $G$,
$M_{R}U_{i}^{T}=U_{i}M_{R}$, $\forall U_{i}\in G$.\ \ 

To be explicit, consider the following three dimensional unitary
representation of $S_{3}=\{I_{i}|i=1\sim6\}$:%
\begin{align*}
I_{1}  &  =\left(
\begin{array}
[c]{ccc}%
1 & 0 & 0\\
0 & 1 & 0\\
0 & 0 & 1
\end{array}
\right)  ,I_{2}=\left(
\begin{array}
[c]{ccc}%
1 & 0 & 0\\
0 & 0 & 1\\
0 & 1 & 0
\end{array}
\right)  ,I_{3}=\left(
\begin{array}
[c]{ccc}%
0 & 1 & 0\\
1 & 0 & 0\\
0 & 0 & 1
\end{array}
\right)  ,\\
I_{4}  &  =\left(
\begin{array}
[c]{ccc}%
0 & 0 & 1\\
0 & 1 & 0\\
1 & 0 & 0
\end{array}
\right)  ,I_{5}=\left(
\begin{array}
[c]{ccc}%
0 & 1 & 0\\
0 & 0 & 1\\
1 & 0 & 0
\end{array}
\right)  ,I_{6}=\left(
\begin{array}
[c]{ccc}%
0 & 0 & 1\\
1 & 0 & 0\\
0 & 1 & 0
\end{array}
\right)  .
\end{align*}

\noindent The four non-trivial subgroups $\{I_{1},I_{2}\}$, $\{I_{1},I_{3}\}$,
$\{I_{1},I_{4}\}$, and $Z_{3}=\{I_{1},I_{5},I_{6}\}$ are all Abelian.
Different from the other three subgroups, the cyclic group $Z_{3}$ is the only
non-trivial invariant subgroup of $S_{3}$. $\{I_{1},I_{5},I_{6}\}$ form a
regular representation of $Z_{3}$. It is straightforward to solve that the
mass matrix $M_{R}$ which satisfies (2) with $U_{i}^{\prime}=U_{i}^{2}$ has
the following form
\begin{equation}
M_{R}=\left(
\begin{array}
[c]{ccc}%
a & b & c\\
b & c & a\\
c & a & b
\end{array}
\right)  =aI_{2}+bI_{3}+cI_{4}. \label{MR}%
\end{equation}
Note that $\{I_{2},I_{3},I_{4}\}$ is a coset of $Z_{3}$ in $S_{3}$.

Before proceeding to the discussion of seesaw mechanism, we would like to
point out another interesting feature of (\ref{MRR}). In fact, before
considering the constraints from symmetry, if one uses a novel mechanism to
generate the most general mass matrix which does not necessarily to be
symmetric, the non-trivial fact is that, with our proposal ($U_{i}^{\prime
}=U_{i}^{2}$), the mass matrix will still be in the form of (\ref{MR}) under
$Z$${}_{3}$ group. This, however, is not true for the case $U_{i}^{^{\prime}%
}=I$ or $U_{i}^{^{\prime}}=U_{i}$. Starting with the most general $M$ with
nine parameters, for the case $U_{i}^{^{\prime}}=I$, one gets
\[
M=\left(
\begin{array}
[c]{ccc}%
a & b & c\\
c & a & b\\
b & c & a
\end{array}
\right)  =aI_{1}+bI_{5}+cI_{6},
\]
while for the case that\ \ $U_{i}^{^{\prime}}=U_{i}$ one gets
\[
M=\left(
\begin{array}
[c]{ccc}%
a & a & a\\
b & b & b\\
c & c & c
\end{array}
\right)  .
\]
But in the symmetric case, these two matrices will become a $\nu$${}_{e}%
-\nu_{\mu}-\nu_{\tau}$ symmetric one and a democratic one, respectively, as
discussed above. That is, the reqirement for $M$ to be symmetric will furthur
reduce the number of free parameters in $M$. On the other hand, unlike the
above two cases, without any assumption on $M_{R}$, starting from our simple
proposal $M_{R}U_{i}^{T}=U_{i}M_{R}$ $\forall U_{i}\in Z_{3}$ and the most
general $M_{R}$ with nine parameters, one still arrives at the unique form of
$M_{R}$ as given by (\ref{MR}).

Assuming that $M_{L}=m_{0}I_{1}$ and $M_{\nu}^{D}=m_{d}I_{1}$ which are
invariant trivially under $Z_{3}$, from (\ref{TypeII}) it can be shown that
the effective neutrino mass can be written as
\begin{equation}
M_{\nu}=mI_{1}+m_{d}^{2}\left(
\begin{array}
[c]{lll}%
B+C & -B & -C\\
-B & A+B & -A\\
-C & -A & C+A
\end{array}
\right)  \label{Mnu}%
\end{equation}

\noindent where $m=m_{0}-m_{d}^{2}(A+B+C),$ and
\begin{equation}
A=\frac{a^{2}-bc}{R},\quad B=\frac{b^{2}-ac}{R},\quad C=\frac{c^{2}-ab}%
{R}\noindent\label{ABC}%
\end{equation}
with $R=a^{3}+b^{3}+c^{3}-3abc$.

This particular form of mass matrix can be diagonalized by the tri-bimaximal
mixing followed by a pure 1-3 rotation. It is worth to mention that any real
symmetric mass matrix which is diagonalized by the tri-bimaximal mixing
followed by a pure 1-3 rotation can always be written in the form of
(\ref{Mnu}).\ \ Therefore what we derive here is a novel way to understand the
phenomenological Majorana neutrino mass matrix with vanishing CP phases that
one can construct from the current neutrino data.

Note that the form of $M_{\nu}$ in (\ref{Mnu}) is coincident with the one in
Friedberg-Lee (FL) model \cite{F-L} in which a new symmetry, i.e. the
invariance of the neutrino mass terms under the transformation
\[
\nu_{e}\longrightarrow\nu_{e}+z,\ \ \nu_{\mu}\longrightarrow\nu_{\mu
}+z,\ \ \nu_{\tau}\longrightarrow\nu_{\tau}+z,
\]
is proposed to explain the observed neutrino mixings. Although more works are
necessary in order to understand the origin of this symmetry and its breaking
mechanism leading to the the first term in the right-hand side of (\ref{Mnu}),
Friedberg and Lee's work provides an illuminating example showing neutrino
physics is a great arena for exploring new physics, which is also what we
pursue here. Although sharing the same motivation to explain neutrino data,
ideas presented in this letter and the physics discussed here are very
different. For example, what Friedberg and Lee discussed are Dirac neutrinos,
but here we consider Majorana neutrinos. Moreover, based on $Z_{3}$ symmetry
and the seesaw mechanism, we provide a simple but new way which can lead to
not only the desired neutrino mass matrix, but also the small neutrino masses.

Before proceed, let's discuss another way to implement $Z_{3}$ symmetry.
Consider the $Z_{3}$ transformation which is realized in the following way
\[
\nu_{1R}\longrightarrow\nu_{1R},\ \ \nu_{2R}\longrightarrow e^{i4\pi/3}%
\nu_{2R},\ \ \nu_{3R}\longrightarrow e^{i2\pi/3}\nu_{3R}%
\]

\noindent and
\[
\phi_{1}\longrightarrow e^{i4\pi/3}\phi_{1},\ \ \phi_{2}\longrightarrow
\phi_{2},\ \ \phi_{3}\longrightarrow e^{i2\pi/3}\phi_{3}
\]

\noindent where $\phi_{i}$ are gauge singlet scalar fields. The invariant
majorana mass terms are
\[
\left(  \overline{{\left(  \nu_{1R}\right)  }^{C}},\overline{{\left(  \nu
_{2R}\right)  }^{C}},\overline{{\left(  \nu_{3R}\right)  }^{C}}\right)
\left(
\begin{array}
[c]{ccc}%
\phi_{2} & \phi_{3} & \phi_{1}\\
\phi_{3} & \phi_{1} & \phi_{2}\\
\phi_{1} & \phi_{2} & \phi_{3}%
\end{array}
\right)  \left(
\begin{array}
[c]{c}%
\nu_{1R}\\
\nu_{2R}\\
\nu_{3R}%
\end{array}
\right)  .
\]

\noindent The VEVs of $\phi_{i}$ will lead to a mass matrix as the one given
in (\ref{MR}). This is equivalent to constructing the following mass term
\[
\overline{{\left(  \nu_{iR}\right)  }^{C}}\phi_{\mathrm{ij}}\nu_{jR}%
\]

\noindent where $\phi$${}_{\mathrm{ij}}$ = $\phi$${}_{(i+j)\ \ \mathrm{mod}
3}$.

We show above that the desired mass matrix can be obtained via $Z_{3}$
symmetry. Although more works are necessary to build a complete model and in
particular, appropriate assignment of the charges of gauge symmetries are
needed, here we concentrate on possible consequences of $Z_{3}$ in the
neutrino sector and assume that other symmetries will not spoil our
discussion. For example, we require any $U(1)$ symmetry or other symmetries,
if exist, will not forbid the required mass terms under discussion.

From (\ref{ABC}), we have $A+B+C=1/(a+b+c)$, and
\begin{equation}
a=\frac{A^{2}-BC}{R^{\prime}},\quad b=\frac{B^{2}-AC}{R^{\prime}}%
,c=\frac{C^{2}-AB}{R^{\prime}}. \label{ABC2abc}%
\end{equation}

\noindent where $R^{\prime}=A^{3}+B^{3}+C^{3}-3ABC$. Now from any set of $A$,
$B$ and $C$ which satisfies experimental data, the corresponding $a$, $b$ and
$c$ can be found by (\ref{ABC2abc}). For heavy right-handed neutrinos,
$m\simeq m_{0}$.

Under tri-bimaximal rotation, we have
\begin{align*}
&  {\left(  U_{0}\right)  }^{T}M_{\nu}U_{0}\\
&  =mI_{1}+m_{d}^{2}\left(
\begin{array}
[c]{lll}%
\frac{3\left(  B+C\right)  }{2} & 0 & \frac{\sqrt{3}}{2}\left(  -B+C\right) \\
0 & 0 & 0\\
\frac{\sqrt{3}}{2}\left(  -B+C\right)  & 0 & \frac{1}{2}\left(  4A+B+C\right)
\end{array}
\right)
\end{align*}

\noindent where
\begin{equation}
U_{0}=\frac{1}{\sqrt{6}}\left(
\begin{array}
[c]{ccc}%
2 & \sqrt{2} & 0\\
-1 & \sqrt{2} & \sqrt{3}\\
-1 & \sqrt{2} & -\sqrt{3}%
\end{array}
\right)  \label{tbm}%
\end{equation}

\noindent is the tri-bimaximal mixing matrix.

The current experimental bound on the matrix element $U_{13}$ given by
$\sin\theta_{13}$ in the standard parameterization is ${\sin}^{2}\theta
_{13}\leq0.040$ at $3\sigma$ C.L. (please see the latest arXiv version of
\cite{Maltoni:2004ei}). This can be satisfied if
\[
{\left(  \frac{B-C}{A-C}\right)  }^{2}={\left(  \frac{b-c}{a-c}\right)  }%
^{2}\ll1.
\]

\noindent Without loss of generality, assume $A>C$. The neutrino masses are
found to be
\begin{align*}
m_{1}  &  \simeq m+\frac{3}{2}m_{d}^{2}(B+C),\\
m_{2}  &  =m,\\
m_{3}  &  \simeq m+2m_{d}^{2}A+\frac{1}{2}m_{d}^{2}(B+C)
\end{align*}

\noindent One can examine that appropriately chosen $a$, $b$ and $c$ can
satisfy the current experimental data. As an example,%
\[%
\begin{array}
[c]{ll}%
m=0.01\quad\mathrm{eV}, & m_{d}=100\quad\mathrm{GeV},\\
a=4.7\times{10}^{14}\quad\mathrm{GeV},\quad & b=5.7\times{10}^{13}%
\quad\mathrm{GeV},\\
c=3.0\times{10}^{13}\quad\mathrm{GeV} &
\end{array}
\]

\noindent will lead to
\[
|U_{13}|=0.022,
\]

\noindent and%
\begin{align*}
&  \Delta m_{21}^{2}=m_{2}^{2}-m_{1}^{2}=7.9\times{10}^{-5}\quad{\mathrm{eV}%
}^{2},\\
&  \Delta m_{31}^{2}=m_{3}^{2}-m_{1}^{2}=2.6\times{10}^{-3}\quad{\mathrm{eV}%
}^{2},
\end{align*}
which are in good agreement with the current neutrino experimental data, i.e.
\begin{align*}
7.1\times{10}^{-5}\quad{\mathrm{eV}}^{2}  &  <\Delta m_{21}^{2}<8.9\times
{10}^{-5}\quad{\mathrm{eV}}^{2}\\
2.0\times{10}^{-3}\quad{\mathrm{eV}}^{2}  &  <\Delta m_{31}^{2}<3.2\times
{10}^{-3}\quad{\mathrm{eV}}^{2}%
\end{align*}
at $3\sigma$ C.L. \cite{Maltoni:2004ei}.

In addition, one can show that this model can account for the case of nearly
degenerate neutrinos. As an example,
\[%
\begin{array}
[c]{ll}%
m=0.25\quad\mathrm{eV}, & m_{d}=15\quad\mathrm{GeV},\\
a=8.93\times{10}^{13}\quad\mathrm{GeV},\quad & b=2.92\times{10}^{12}%
\quad\mathrm{GeV},\\
c=9.01\times{10}^{11}\quad\mathrm{GeV} &
\end{array}
\]
will lead to the same squared mass differences as given above and
$|U_{13}|=0.008$.

In conclusion, we show that $Z_{3}$ symmetry can lead to observed neutrino
mixing. We find that if one requires $M_{R}U_{i}^{T}=U_{i}M_{R}$, $\forall
U_{i}\in Z_{3}$, $M_{R}$ must be in a cyclic permuted form, as shown in
(\ref{MR}). This will lead to tri-bimaximal mixing followed by an additional
1-3 rotation. Another way is based on the invariance of the mass terms under
$Z_{3}$ transformations, similar to the usual $Z_{2}$ R-parity
transformations. In the seesaw framework, this will lead to a possible
explanation to both the smallness of neutrino masses and the origin of the
neutrino mixing. It can be easily shown that $\theta_{13}=0$ requires $b=c$ in
(\ref{MR}), i.e., the $\nu_{\mu}-\nu_{\tau}$ symmetry. Therefore, from
naturalness principle, the smallness of $\theta$${}_{13}$ is presumably
protected by the symmetry. However, what we ignored here is the $\nu{}_{\mu
}-\nu_{\tau}$ symmetry breaking mechanism leading to the smallness of
$\sin\theta_{13}$, and other possible phenomena including lepton flavor
violations (LFV), which is worth further studies in the future.

\section*{Acknowledgements}

This work was supported in part by the key projects of Chinese Academy of
Sciences, the National Science Foundation of China (NSFC) under the grant
10475105, 10491306. This research was also supported in part by the National
Science Foundation under Grant No. PHY99-07949. B.H.'s work was also supported
by the National Science Foundation of China (NSFC) under the grant 10505011
and 10663001, Jiangxi Provincial Department of Education under the Science and
Technology Research Project grant 2006-17 and the Program for Innovative
Research Team of Nanchang University.


\begin{thebibliography}{99}                                                                                               %


\bibitem {MohapatraReview}R. N. Mohapatra and A. Y. Smirnov, arXiv:hep-ph/0603118.

\bibitem {ValleReview}J. W. F. Valle, arXiv:hep-ph/0608101.

\bibitem {Strumia:2006db}A. Strumia and F. Vissani, arXiv:hep-ph/0606054.

\bibitem {HPS}P. F. Harrison, D. H. Perkins and W. G. Scott, Phys. Lett.
\textbf{B 530}, 167 (2002).

\bibitem {seesaw2}G. Lazarides, Q. Shafi and C. Wetterich, Nucl. Phys.
\textbf{B181}, 287 (1981); R. N. Mohapatra and G. Senjanovi\'{c}, Phys. Rev.
\textbf{D 23}, 165 (1981).

\bibitem {HabaS3}N. Haba and K. Yoshioka, Nucl. Phys. \textbf{B 739}, 254 (2006).

\bibitem {HagedornS4}C. Hagedorn, M. Lindner and R. N. Mohapatra, JHEP
\textbf{0606}, 042 (2006).

\bibitem {MaA4}E. Ma, H. Sawanaka and M. Tanimoto, Phys. Lett. B \textbf{641},
301 (2006).

\bibitem {F-L}R. Friedberg and T. D. Lee, arXiv:hep-ph/0606071.

\bibitem {tri-bi}L. Wolfenstein, Phys. Rev. \textbf{D 18}, 958 (1978); P. F.
Harrison and W. G. Scott, Phys. Lett. \textbf{B 535}, 163 (2002); Z.~z.~Xing,
Phys.\ Lett.\ \textbf{B 533}, 85 (2002);

\bibitem {Maltoni:2004ei}M. Maltoni, T. Schwetz, M. A. Tortola and J. W. F.
Valle, New J. Phys. \textbf{6}, 122 (2004) [arXiv:hep-ph/0405172].
\end{thebibliography}
\end{document}